\documentclass[conference]{IEEEtran}

\usepackage{cite}
\usepackage{microtype}
\usepackage{amsmath}
\usepackage{amssymb}
\usepackage{braket}
\usepackage{hyperref}
\hypersetup{hidelinks=true,draft}
\usepackage{tikz}
\usepackage{circuitikz}
\usetikzlibrary{patterns,chains,fit,positioning,decorations.markings}
\usepackage{array}
\usepackage{url}
\usepackage{soul}
\begin{document}
\bstctlcite{IEEEexample:BSTcontrol}

\title{A comparison between quantum and classical noise radar sources}

\IEEEoverridecommandlockouts

\author{\IEEEauthorblockN{Robert Jonsson}
	\IEEEauthorblockA{$^1$Department of Microtechnology\\ and 
	Nanoscience\\Chalmers University of Technology\\
		$^2$Radar Solutions, Saab AB\\
		G\"oteborg, Sweden\\
		Email: robejons@chalmers.se}\\
	\IEEEauthorblockN{Anders Str\"om}
	\IEEEauthorblockA{Radar Solutions, Saab AB\\
		G\"oteborg, Sweden}
	\and
	\IEEEauthorblockN{Roberto Di Candia}
	\IEEEauthorblockA{Department of Communications\\ and Networking\\Aalto 
	University\\
		Helsinki, Finland\\
		Email: roberto.dicandia@aalto.fi}\\[1em]
	\IEEEauthorblockN{G\"oran Johansson}
	\IEEEauthorblockA{Department of Microtechnology\\ and Nanoscience\\Chalmers 
	University of Technology\\
		G\"oteborg, Sweden}
	\and
	\IEEEauthorblockN{Martin Ankel}
	\IEEEauthorblockA{$^1$Department of Microtechnology\\ and 
	Nanoscience\\Chalmers University of Technology\\
		$^2$Radar Solutions,	Saab AB\\
		G\"oteborg, Sweden}
}

\maketitle

\begin{abstract} We compare the performance of a quantum radar based on 
two-mode 
squeezed
states with a classical radar 
system based on correlated thermal noise. 
With a constraint of equal number of photons $N_S$ transmitted to 
probe the environment, we find that the quantum setup exhibits an advantage 
with respect to its classical counterpart of $\sqrt{2}$ in the cross-mode 
correlations. Amplification of the signal and the idler is considered at 
different stages of the protocol, showing that no quantum advantage is 
achievable when a large-enough gain is applied, even when quantum-limited 
amplifiers are available. We also characterize the minimal type-II error 
probability decay, given a constraint on the type-I error probability, and find 
that the optimal decay rate of the type-II error probability in the quantum 
setup is  $\ln(1+1/N_S)$ larger than the optimal classical setup, in the 
$N_S\ll1$ regime.  In addition, we consider the Receiver Operating 
Characteristic (ROC) curves for the scenario when the idler and the received 
signal are measured separately, showing that no quantum advantage is present in 
this case. Our work characterizes the trade-off between quantum correlations 
and noise in quantum radar systems.
\end{abstract}

\IEEEpeerreviewmaketitle

\section{Introduction}
The quantum Illumination (QI) 
protocol~\cite{lloyd2008enhanced,tan2008quantum,Lopaeva2013,Zhang2013,Zhang2015,
 Quntao2017, Sanz2017} uses entanglement as a resource to improve the detection 
of a low-reflectivity object embedded in a bright environment. The protocol was 
first developed for a single photon source~\cite{lloyd2008enhanced}, and it was 
then extended to general bosonic quantum states and thermal bosonic 
channels~\cite{tan2008quantum}. Here, a $6$~dB advantage in the effective 
signal-to-noise ratio (SNR) is achievable when using two-mode squeezed states 
instead of coherent states. This gain has been recently shown to be optimal, 
and reachable {\it exclusively} in the regime of low transmitting power per 
mode~\cite{DiCandia2020,Nair2020}. The QI protocol has possible applications in 
the spectrum below the Terahertz band, as here the environmental noise is 
naturally bright. In particular, microwave quantum technology has been very 
well developed in the last decades~\cite{Nori2017}, paving the way of 
implementing these ideas for building a first prototype of quantum radar. 

The possible benefits of a quantum radar system are generally understood to be 
situational. In an adversarial  scenario, it is beneficial for a radar system 
to be able to
operate while minimizing the power output, in order to reduce the probability 
for the 
transmitted signals 
to be detected. This
property is commonly referred to as Low Probability of Intercept (LPI), and it 
is a common measure to limit the ability of the enemy to localize and discover 
the radar. The low signal levels required for QI are in principle excellent for 
acquiring good LPI properties. However, there are several challenges to face in 
order to achieve this goal.
A first proposal for implementing a microwave QI protocol was advanced in 
Ref.~\cite{barzanjeh2015microwave}. The protocol relies on an efficient 
microwave-optical interface for the idler storage and the measurement stage. 
This technology is promising for this and other applications, however it is 
still in its infancy. Furthermore, the signal generation requires cryogenic 
technology, which must be interfaced with a room-temperature environment. 
Recently, a number of QI-related experiments have been carried out in the 
microwave regime~\cite{chang2019quantum,barzanjeh2019experimental}, showing 
that some correlations of an entangled signal-idler system are preserved after 
the signal is sent out of the dilution refrigerator. While these results are a 
good benchmark for future QI experiments, they strictly rely on the 
amplification of the signal and idler. This has been shown to rule out any 
quantum advantage with respect to an optimal classical 
reference~\cite{shapiro2019quantum}. 

In this work, we discuss the role of quantum correlations and amplification in 
the QI protocol, providing a comparative analysis of quantum and classical 
noise radars in different scenarios. Noise radar is an old concept that 
operates by probing the environment with a noisy signal and cross-correlating 
the returns with a retained copy of the transmitted 
signal~\cite{cooper1967random}.
A \textit{quantum} noise radar  operates  similarly to 
its 
conventional counterpart, but differs in the use of a two-mode entangled 
state as noise source~\cite{chang2019quantum,barzanjeh2019experimental}. An 
advantage of the quantum noise radar over the 
classical counterpart can be declared 
if stronger correlations can be achieved, when both systems 
illuminate the environment with equal power. In the microwave regime, the 
two-mode squeezed state used for noise 
correlations can be generated with 
superconducting circuits with a Josephson Parametric Amplifier (JPA) at 
$T\simeq 20$~mK~\cite{Flurin2012, Menzel2012}. On the one hand, using quantum 
correlated signals generated by a JPA enhances the signal-to-noise ratio in the 
low transmitting power per mode regime. On the other hand, Josephson parametric 
circuits are able to generate correlated and entangled signals with large 
bandwidth~\cite{Wilson2011,johansson2013nonclassical,schneider2020observation}. 
This allows, in principle, a system to operate in the low power-per-mode 
regime, where quantum radars show fully their advantage.
Here, we analyze the performance of a JPA-based noise radar in different 
scenarios which include different sources of noise. Our analysis shows that any 
quantum advantage is destroyed by the unavoidable noise added when amplifying 
either the signal or the idler. We also show when the idler and signal are 
measured separately, the entanglement initially present in the signal-idler 
system is not properly exploited, and no quantum advantage can be retained. The 
latter happens even without amplifying the signal or the idler. Our work 
complements the analysis done in Ref.~\cite{shapiro2019quantum} with the 
explicit calculations of the cross-correlation coefficients and the optimal 
asymptotic ROC performance in the microwave regime.

\section{Theory}\label{sec:theory}

In this section, we introduce the models for the quantum and classical systems, 
within the quantum mechanical description. In this step, we emulate 
Refs.~\cite{luong2019receiver,shapiro2019quantum}, where the classical and 
quantum noise radar were first studied. In all expressions, we
assume the natural units ($\hbar =1$, $k_\text{B}=1$).

\subsection{Quantum preliminaries}\label{ssec:quantum}

A single, narrowband mode of the electric field,  at microwave frequency $f$, 
is defined with an 
operator (in suitable units) as
$\hat{E} =	\hat{q}\cos{2\pi f t} + \hat{p}\sin{2\pi ft},$
where $\hat{q}$ and $\hat{p}$ are the in-phase and quadrature operators, 
respectively. The quadratures are related to the bosonic annihilation 
($\hat{a}$) and creation ($\hat{a}^\dagger$)
operators as 
$\hat{q}=(\hat{a}^\dagger{+}\hat{a})/\sqrt{2}$ and
$\hat{p}=\mathrm{i}(\hat{a}^\dagger{-}\hat{a})/\sqrt{2}$, where 
$[\hat{a},\hat{a}^\dagger]=\hat{\mathbb{I}}$.
The commutation relation
$[\hat{q},\hat{p}]=\mathrm{i}\hat{\mathbb{I}}$ implies that the quadratures can 
not be measured 
simultaneously with arbitrary precision, due to the Heisenberg uncertainty 
relation. In the following, we represent the quadratures of the two-modes of 
the electric field by the 
vector $\hat{X}=(\hat{q}_S,\hat{p}_S,\hat{q}_I,\hat{p}_I)^T$,
where the indices $S$ 
and 
$I$ refer 
to the signal and idler mode, respectively. These 
mode designations are used interchangeably for both the quantum and classical 
system.

\subsubsection{Classically-correlated thermal noise}

\begin{figure}
	\begin{tikzpicture}[decoration={
	markings,
	mark=at position 0.5 with {\arrow{>}}}
]

\node at (0,0) {};
\node [draw] at (2.5,0.5) (T1) {$T_1$};
\node [draw] at (2.5,3.5) (T0) {$T_0$};

\draw [postaction={decorate}] (T0.south east) -- ($(T0.south 
east)+(1.3,-1.3)$);
\draw [postaction={decorate}] (T1.north east) -- ($(T1.north 
east)+(1.3,1.3)$);
\draw [postaction={decorate}] ($(T1.north east)+(1.3,1.3)$) -- ($(T1.north 
east)+(2.5,2.5)$);
\draw [postaction={decorate}] ($(T0.south east)+(1.3,-1.3)$) -- ($(T0.south 
east)+(2.5,-2.5)$);

\draw [pattern=north west lines, pattern color=black] 
(3.2,1.975) rectangle (4.8,2.025) ;
\node at (5.3,2) {$(\xi,\varphi)$};

\node at (3.5,2.9) {$\hat{a}_0$};
\node at (3.5,1.1) {$\hat{a}_1$};
\node at (5.7,3.5) {$\hat{a}_S^{(C)}$};
\node at (5.7,0.5) {$\hat{a}_I^{(C)}$};
\end{tikzpicture}
	\caption{{\bf Preparation of classically-correlated thermal noise.} A 
	beamsplitter with reflection coefficient $\xi$ and phase turning angle 
	$\varphi$ generates the signal and idler modes from the modes $\hat{a}_0$ 
	and $\hat{a}_1$.  These modes are in a thermal state with $T_0$ and $ T_1$ 
	effective noise temperatures, respectively. The output modes are correlated 
	provided that $T_0\not= T_1$.}\label{fig:beamsplitter}
\end{figure}
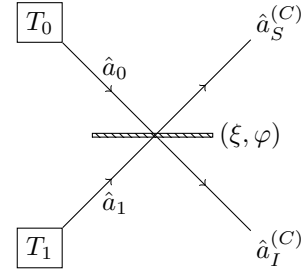

The classically-correlated noise (CCN) system uses two sources  of thermal 
noise, $\hat a_0$ and $\hat a_1$, at temperatures $T_0$ and $T_1$, 
respectively. In general, the quantum state of a thermal noise mode at 
temperature $T$ can be represented by the density operator
\begin{equation}
\mathbf{{\rho}}_{th} = 
\sum_{n=0}^{\infty}\frac{N^n}{\left(N+1\right)^{n+1}}\ket{n}\bra{n},
\end{equation}
where the average number\footnote{All variables using the symbol $N$ refer 
	to mode quanta and should 
	not be confused with 
	the noise \textit{figure} of a microwave component, which often shares the 
	same 
	symbol.} of photons is defined by 
the thermal equilibrium Bose-Einstein statistics at temperature $T$, i.e., 
$N=\left[\exp\left(2\pi 
f/T\right)-1\right]^{-1}$. In the following, we will refer as $N_0$ ($N_1$) the 
average number of photons for the mode $\hat a_0$ ($\hat a_1$). The thermal 
modes $\hat a_0$ and $\hat a_1$ pass
through a 
beamsplitter, as shown in Fig.~\ref{fig:beamsplitter}. This generates a signal 
mode $\hat{a}_S^{(C)}$ and an idler mode $\hat{a}_I^{(C)}$, related to the 
inputs as~\cite{loudon2000quantum} 
\begin{equation}
\begin{pmatrix}
\hat{a}_S^{(C)}\\\hat{a}_I^{(C)}
\end{pmatrix} = \begin{pmatrix}
\sqrt{\xi} & \sqrt{1-\xi}\mathrm{e}^{\mathrm{i}\varphi} 
\\-\sqrt{1-\xi}\mathrm{e}^{-\mathrm{i}\varphi} & \sqrt{\xi} 
\end{pmatrix}\begin{pmatrix}
\hat{a}_0 \\ \hat{a}_1
\end{pmatrix}\label{eq:beamsplitter}.
\end{equation}
Here, $\xi\in(0,1)$ is the reflection coefficient 
and $\varphi$ is the phase turning angle of the 
beamsplitter, in the following set to be zero. 
One can think of this process as a noise signal, generated by a thermal source 
at temperature $T_0$, sent as input of a power divider placed in an environment 
at temperature $T_1$. The output modes $\hat{a}_S^{(C)}$ and $\hat{a}_I^{(C)}$ 
are in a thermal state with $\xi N_0+(1-\xi)N_1$ and $\xi N_1+(1-\xi)N_0$ 
average number of photons, respectively. If $T_1\not=T_0$, or, equivalently, 
$N_1\not=N_0$, then the outputs are classically-correlated, regardless of the 
value of $\xi$.

\subsubsection{Entangled thermal noise}

A Two-Mode Squeezed
Vacuum (TMSV) state $\ket{\psi}_{\rm TMSV}$ is represented in the 
Fock basis as 
\begin{equation}
\ket{\psi}_{\rm 
TMSV}=\sum_{n=0}^{\infty}\sqrt{\frac{N_S^n}{\left(N_S+1\right)^{n+1}}}\ket{n}_S\ket{n}_I,
\end{equation}
where $N_S$ is the average number of photons in both 
the 
signal and idler mode.
A TMSV state is closely related to a classically-correlated thermal noise, as 
also here both signal and idler photons are Bose-Einstein distributed. However, 
as we will see, the resulting correlations in the low signal-power regime are 
{\it stronger} for the TSMV states.

\subsection{Covariance and correlation matrices}

As the states  considered here are Gaussian, their statistics is entirely 
determined by 
the first and second order quadrature moments. For zero-mean states, i.e., 
when  $\braket{\hat{X}_i}=0$ for all $i$, the states are 
characterized entirely by 
the covariance matrix $\mathbf{\Sigma}$, with elements 
\begin{equation}\label{covariance}
\mathbf{\Sigma}_{i,j} = 
\frac{1}{2}\braket{\hat{X}_i\hat{X}_j^\dagger+\hat{X}_j\hat{X}_i^\dagger}-\braket{\hat{X}_i}\braket{\hat{X}_j^\dagger}.
\end{equation}
This is the case for both the classical and the entangled thermal noise states. 
Similarly, one can introduce the correlation coefficient matrix  $\mathbf{R}$, 
whose elements are 
\begin{equation}
\mathbf{R}_{i,j} = 
\frac{\Sigma_{i,j}}{\sqrt{\Sigma_{i,i}}\sqrt{\Sigma_{j,j}}}\in[-1,1].\label{eq:pearson}
\end{equation}
These coefficients, also referred to as 
\textit{Pearson's correlation coefficients}, characterize the linear 
dependence between the quadratures $\hat{X}_i$ and $\hat{X}_j$.

\subsection{Quantum relative entropy}\label{relative_entropy}
The quantum relative entropy defines an information measure between two quantum 
states. It is defined as 
\begin{equation}
    D(\rho_1||\rho_0)=\text{Tr}\,\rho_1(\ln \rho_1-\ln\rho_0),
\end{equation}
for two density matrices $\rho_0$ and $\rho_1$. This quantity is related to the 
performance in the asymmetric binary hypothesis testing via the quantum Stein's 
lemma. The task is to discriminate between $M$ copies of $\rho_0$ and $M$ 
copies of $\rho_1$,  given a bound on the type-I error probability 
(\textit{probability of false alarm}, $P_{Fa}$) of $\varepsilon\in(0,1)$. In 
this discrimination, the maximum type-II error probability (\textit{probability 
of miss}, $P_M$) exponent is
\begin{equation}\label{Steins}\small
    -\frac{\ln P_M}{M} = D(\rho_1 || \rho_0)+\sqrt{\frac{V(\rho_1 || 
    \rho_0)}{M}}\Phi^{-1}(\varepsilon)+\mathcal{O}\left(\frac{\ln M}{M}\right),
\end{equation}
where 
$V(\rho_1||\rho_0)=\text{Tr}\,\rho_1[\ln\rho_1-\ln\rho_0-D(\rho_1||\rho_0)]^2$ 
is the quantum relative entropy variance and $\Phi^{-1}$ is the inverse 
cumulative normal distribution~\cite{Wilde2017}. In this work, we rely on 
quantum relative entropy computations and its variance in order to quantify the 
performance in the asymptotic setting, i.e., when $M\gg1$. This is in contrast 
to the original treatment based on the Chernoff bound~\cite{tan2008quantum}, 
which provides an estimation of the average error probability when the prior 
probabilities of target absence or presence are equal. In a typical radar 
scenario, the prior probabilities are not the same, and may be even unknown.

\section{Noise radar operation}\label{sec:radar}

In this section, we analyze the performance of the classical and quantum noise 
correlated radars, based on the states defined in the previous section. 

\subsection{Probing the environment}

\begin{figure}
	\begin{tikzpicture}[decoration={
	markings,
	mark=at position 0.5 with {\arrow{>}}}
]
\node [draw] at (0,0) (P) {Preparation};
\node [draw] at (7,0) (D) {Detection};
\draw[black] ([yshift=4]P.east)--++(0:.2) node[antenna] (A1) {};
\coordinate (AP1) at ($(A1)+(0.35,2.1)$);
\draw[black] ([yshift=4]D.west)--++(0:-.2) node[antenna] (A2) {};
\coordinate (AP2) at ($(A2)+(-0.35,2.1)$);

\node (R) at (3.5,3.9) {};
\node (R1) at (3.5,4.2) {};
\draw [pattern=north west lines, pattern color=black] 
(2.5,4.29) rectangle (4.5,4.34) ;
\draw [postaction={decorate}] ([yshift=-4]P.east) -- ([yshift=-4]D.west);
\draw [postaction={decorate}] (AP1) -- ($(AP1)+(2,2)$);
\draw [postaction={decorate}] ($(AP2)+(-2,2)$) -- (AP2);
\draw [postaction={decorate}] ($(AP2)+(-3,3)$)--($(AP2)+(-2,2)$);
\draw [postaction={decorate}] ($(AP1)+(2,2)$)--($(AP1)+(3,3)$);
\node at (3.5,0.1) {$\hat{a}_I$};
\node at (2.2,3.4) {$\hat{a}_S$};
\node at (4.9,3.4) {$\hat{a}_R$};
\node at (3.15,5.1) {$\hat{a}_B$};
\node at (4.7,4.3) {$\eta$};

\end{tikzpicture}
	\caption{{\bf Scheme of the quantum and classical noise radar systems.} The
		system probes a region of space with a signal $\hat{a}_S$ to detect a 
		possible object, modelled as a channel
		with reflectivity parameter $\eta$. The returned signal $\hat a_R$ is 
		then used for detecting correlations with the idler mode $\hat{a}_I$, 
		which has been retained in the lab.}\label{fig:overview}
\end{figure}
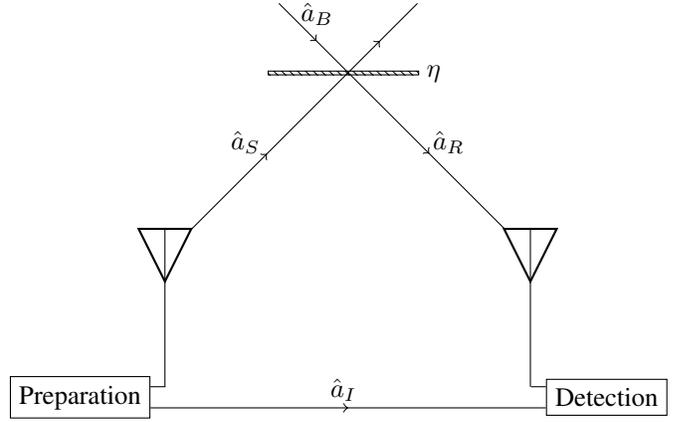

The signal mode is 
transmitted to probe the environment where an object (\textit{target}) may be 
present or absent. This process is modelled as a channel with  reflection 
coefficient 
$\eta$ that is non-zero and small when the target is present ($0<\eta\ll 1$) 
and zero when the target is absent ($\eta =0$), see 
Fig.~\ref{fig:overview}.  
Here, $\eta$ can be interpreted as the ratio between transmitted 
power and received power, including the effects of atmospheric attenuation, the 
antenna 
gain and the target radar cross section. We use a beamsplitter model to take 
into account the environmental losses. In other words, the returned mode $\hat 
a_R$ is given by
\begin{equation}\label{return}
\hat{a}_R = 
\sqrt{\eta}~\hat{a}_S\mathrm{e}^{-\mathrm{i}\theta}+\sqrt{1-\eta}~\hat{a}_B,
\end{equation}
where $\hat{a}_B$ is a bright background noise mode with 
$\braket{\hat{a}_B^\dagger\hat{a}_B} = N_B$ average power per mode, and where 
$\theta$ is a phase shift relative to the idler. In the $1-10$~GHz regime, 
where the technology is advanced enough to apply these ideas in the quantum 
regime, we have that $N_B\simeq 10^3$, which is assumed for numerical 
computations. For the current calculation,
the reflection coefficient is assumed to be non-fluctuating. We also assume 
$\braket{\hat{a}_I\hat{a}_B}=0$, i.e., the returned signal preserves some 
correlations with the idler mode only if the object is present. This allows us 
to define a correlation detector able to detect the presence or absence of the 
object.

\subsection{Cross-correlation coefficient}
The covariance matrix of the system composed of the received signal and the 
idler is easily computable using Eq.~\eqref{covariance} and Eq.~\eqref{return}. 
For both the classical and the quantum noise radars considered here, applying 
Eq.~\eqref{eq:pearson} gives us the correlation matrix
\begin{equation}
\mathbf{R} = \begin{pmatrix}
\mathbf{I} & \kappa \mathbf{D}(\theta)  \\ \kappa \mathbf{D}(\theta)^T & 
\mathbf{I}
\end{pmatrix},\label{eq:corrMat}
\end{equation}
where $0\leq\kappa\leq 1$ is the amplitude of the cross-correlation 
coefficient, 
 and  
$\mathbf{D}$ is a matrix with determinant $\left|\mathbf{D}\right|=\pm1$. 
The cross-correlation coefficient for the entangled TMSV source can be derived 
directly from the definition,
\begin{align}
    	\kappa_\text{TMSV} &= \frac{\sqrt{\eta 
	N_S(N_S+1)}}{\sqrt{N_R+\frac{1}{2}}\sqrt{N_S+\frac{1}{2}}},\label{eq:kappaQRI}
\end{align}
with $N_R =\eta 
N_S+(1-\eta)N_B$. 
For a fair comparison between the quantum and classical systems, we introduce a 
constraint 
on the transmitted power of the signal modes, i.e., we set 
\begin{equation}
N_S = \xi N_0+(1-\xi)N_1.\label{eq:constraint0}
\end{equation}
This constraint can be interpreted as giving both systems equal LPI properties. 
Eq.~\eqref{eq:constraint0} yields an expression for the classical 
cross-correlation amplitude as
\begin{equation}
\kappa_\text{CCN}=\frac{\sqrt{\eta}(N_S-N_1)}{\sqrt{N_R+\frac{1}{2}}\sqrt{N_S-N_1+\frac{\xi}{1-\xi}\left(N_1+\frac{1}{2}\right)}},\label{eq:kappaCres}
\end{equation}
where $N_S>N_1$ follows directly from Eq.~\eqref{eq:constraint0} and the 
assumption $N_0>N_1$.
This quantity is maximal in the $N_1\ll1$ regime, where 
Eq.~\eqref{eq:constraint0} reduces to $N_S = \xi N_0$. We assume $N_1 \ll1$, 
which  corresponds to classically-correlated thermal noise generated in a 
noise-free environment. At microwave frequencies this is achievable at mK 
temperatures. Eq.~\eqref{eq:kappaCres}, for given noise transmitting power, 
defines the correlations of a class of classical noise radars, labeled by the 
beamsplitter parameter $\xi$.

\subsection{Quantum advantage}
\begin{figure}
	\includegraphics[width=\linewidth]{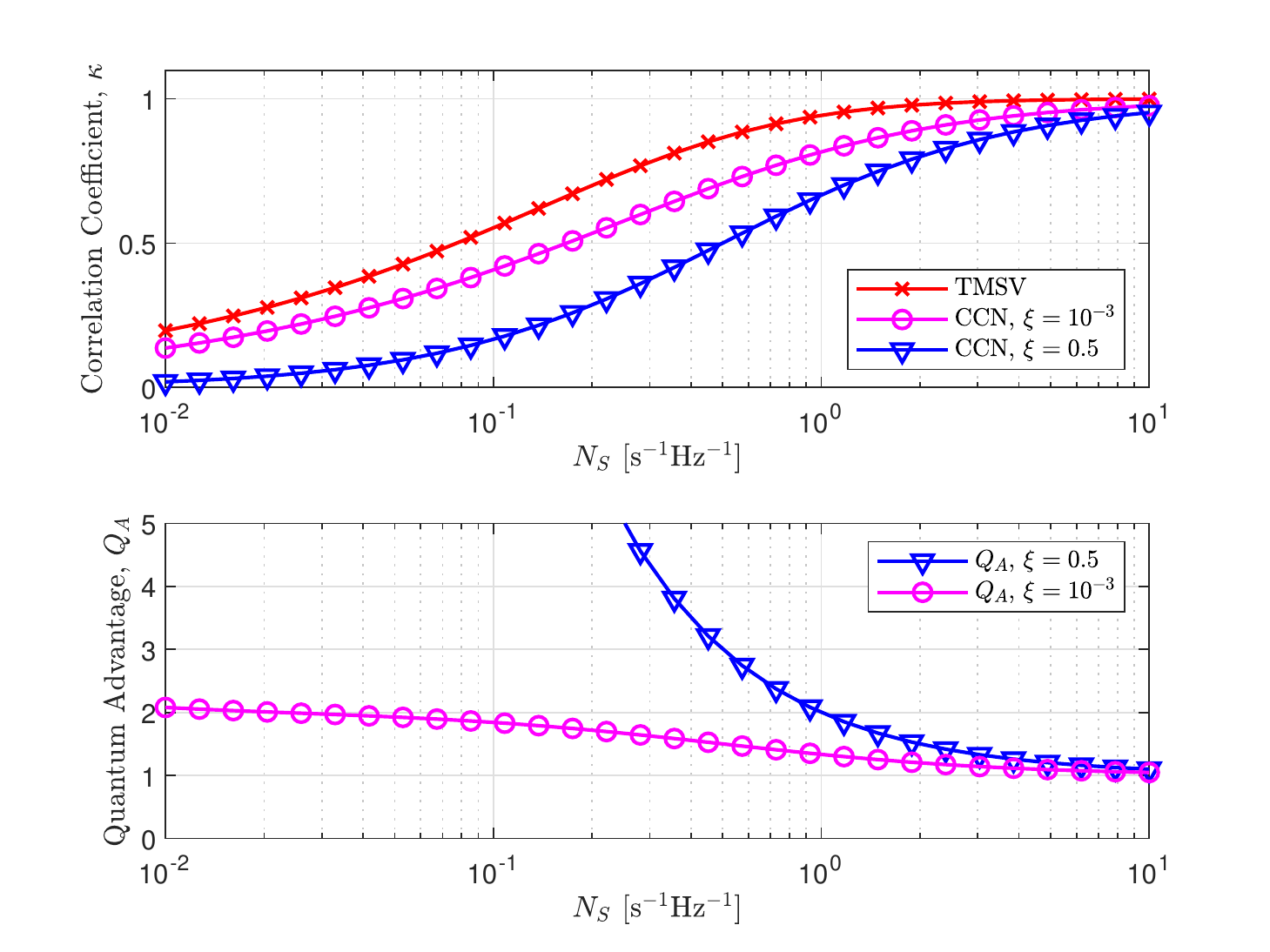}
	\caption{{\bf Performance of the quantum and classical noise radars in 
	terms of the cross-correlation coefficient.} (Upper) The cross-correlation 
	coefficients for the quantum and classical noise radars, as functions of 
	the transmitted quanta, rescaled such that $\sqrt{\eta}=1$. (Lower) The 
	quantum advantage for the cases considered 
		above. In both plots, 
		we have considered the $\xi=10^{-3}$ and $\xi=0.5$ settings. The first 
		setting achieves a close to optimal cross-correlation coefficient. The 
		second setting is suboptimal, and it has been used as classical 
		reference in recent microwave illumination 
		experiments~\cite{chang2019quantum,barzanjeh2019experimental}. 
		}\label{fig:cross-corr}
\end{figure}

It is easy to see that the quantity $\kappa^2$ is linearly proportional to the 
\textit{effective} SNR in the likelihood-ratio tests. A larger value of 
$\kappa^2$ means a stronger discrimination power.  Therefore, we define a 
figure of merit $Q_A$, quantifying the advantage of the quantum over the 
classical noise radar, as
\begin{align}\label{eq:advantage}
Q_A &\equiv \frac{\kappa_\text{TMSV}^2}{\kappa_\text{CCN}^2} =  
\frac{N_S+1}{N_S+\frac{1}{2}}\left[1+\frac{\xi}{2N_S(1-\xi)}\right],
 \end{align}
which can be evaluated for different values of the free parameter $\xi$.
Restricting the constraint  to equal power in both the signal and idler 
mode is equivalent to applying a 50-50
beamsplitter to the thermal noise source in Eq.~\eqref{eq:beamsplitter}, or, in 
other words,  it corresponds to setting $\xi~=~1/2$. This gives
\begin{equation}
Q_{A}(\xi=1/2) = 1+\frac{1}{N_S},
\end{equation}
which is unbounded for $N_S\rightarrow 0$. This setting as been used as 
benchmarking in the recent microwave quantum illumination 
experiments~\cite{chang2019quantum,barzanjeh2019experimental}. However, this 
choice of $\xi$ is not optimal, leading to an overestimation of the quantum 
radar advantage\footnote{This criticism was already raised by J. H. Shapiro in 
Ref.~\cite{shapiro2019quantum}.}. A strongly 
\textit{asymmetric} beamsplitter must be applied to a very bright noise 
source ($N_0\gg1$) in order to 
maximize the correlations in the classical case, while maintaining the 
constraint on transmitted power $N_S=\xi N_0$.
It can be seen  in Eq.~\eqref{eq:advantage}, that $Q_A$ is maximized in the 
$\xi/(1-\xi)\ll N_S$ limit. In this setting, the CCN idler has a much better 
SNR than in the symmetric configuration, and we get
\begin{equation}
Q_{A}(\xi/(1-\xi)\ll N_S)\approx
2\left[1-\frac{N_S}{2N_S+1}\right].
\end{equation}
In the low transmitting power limit we have that  $\lim\limits_{N_S\rightarrow 
0}Q_{A}=2$, i.e., a $\sqrt{2}$ advantage in the correlation coefficient. 
In  Fig.~\ref{fig:cross-corr} we show the functional behaviour of the 
correlation coefficients for the entangled and classically-correlated thermal
noise source, depending on the transmitting power $N_S$. The quantum advantage 
decays slowly with increasing $N_S$, keeping an advantage also for finite 
$N_S$, until virtually disappearing for $N_S\simeq 10$.
Note that in the limit where the advantage is 
maximized, 
the modes also become uncorrelated ($\kappa\rightarrow0$) in both systems. 
A signal-idler system with a large operating bandwidth is needed to compensate 
the low amount of correlations per mode in the $N_S\ll1$ limit. 

Here, the phase-shift of the signal mode due to the propagating path has been 
set to zero. In other words, we work in a rotated frame, where the inter-mode 
phase angle $\theta$ is applied only to the idler mode, i.e., $\hat{a}_I 
\mapsto 
\hat{a}_I\mathrm{e}^{-\mathrm{i}\theta}$. 
Note that, in general, the phase 
$\theta$ is unknown. The original QI protocol assumes the knowledge of the 
inter-mode phase angle $\theta$. 
In this case, the complex conjugate receiver defined in Ref.~\cite{Guha2009} 
saturates the quantum advantage given in Eq.~\eqref{eq:advantage} with a 
likelihood-ratio test~\cite{Zhuang2017neyman}. If $\theta$ is unknown, then one 
can define an adaptive strategy where $\mathcal{O}(\sqrt{M})$ copies are used 
to get a rough estimate of $\theta$, and then $M-\mathcal{O}(\sqrt{M})$ are 
used to perform the discrimination protocol in the optimal reference frame 
maximizing the Fisher information~\cite{Sanz2017}. This strategy shows the same 
asymptotic performance as in the case of known $\theta$.

\subsection{The effect of amplification}
In the following, we consider three Gaussian amplifying schemes. We show how 
the quantum advantage defined in terms the cross-correlation coefficients 
rapidly disappears when an amplification is involved at any stage of the 
protocol.

\subsubsection{Amplification before transmitting the signal 
mode}\label{ssec:SAmp}

If we amplify the signal mode before transmission to the 
environment, the transmitted mode is
$\hat{a}_S'=\sqrt{G_S}~\hat{a}_S+\sqrt{G_S-1}~\hat{a}_{G_S}^\dagger,$
where $G_S\geq1$ is the gain and where $\braket{\hat{a}_{G_S}^\dagger 
\hat{a}_{G_S}}=N_{G_S}$ is the added noise. The classical and quantum 
covariance matrix transforms identically.
Correspondingly, the 
advantage in Eq.~\eqref{eq:advantage} is actually
\textit{independent} of the signal amplification. However, although $G_S$ 
counteracts the 
loss due to $\eta$, the added noise   
limits the absolute correlations that can be achieved. In addition, in the weak 
signal regime where the quantum advantage is relevant, the effective SNR is low 
even before 
amplification. Amplifying a signal with low SNR  is trivially outperformed by 
using 
a stronger signal with better SNR to begin with, as is the case of a 
classically-correlated thermal noise with $G_SN_S+(G_S-1)(N_{G_S}+1)$ average 
signal photons. In addition, any sufficiently strong amplification applied at 
this stage would weaken LPI properties, and possibly move the system to a 
regime where another  protocol outperforms correlation detection.

\subsubsection{Amplification before measurement}

These amplification schemes describe the pre-amplification of the modes in 
heterodyne quadratures measurement.

With $G_R\geq1$ amplifying the returned mode, as
$\hat{a}_R'=\sqrt{G_R}~\hat{a}_R+\sqrt{G_R-1}~\hat{a}_{G_R}^\dagger,$
the covariance matrix again transforms identically  for the quantum and 
classical  systems, 
and any added noise cancels in the advantage merit. However, as can be 
intuitively 
understood, 
amplification at the receiver end can not increase correlations with the idler 
reference. Indeed, the added amplifier noise 
reduces correlations, 
but equally in both systems. 

Instead, applying the amplification $G_I\geq1$ to the idler, as
$\hat{a}_I'=\sqrt{G_I}~\hat{a}_I+\sqrt{G_I-1}~\hat{a}_{G_I}^\dagger,$
is actually 
detrimental to the 
advantage of Eq.~\eqref{eq:advantage}. 
In the limit of strong amplification ($G_I\gg1$), the advantage merit is
\begin{equation}
    Q_A = 
    \frac{N_S+1}{N_S+N_{G_{I}}+1}\left[1+\frac{\xi(N_{G_I}+1)}{N_S(1-\xi)}\right],
\end{equation}
where $\braket{\hat{a}_{G_I}^\dagger\hat{a}_{G_I}}=N_{G_I}$ is the added noise.
We consider the ideal case of quantum limited amplifier noise 
($N_{G_I}\rightarrow0$), which simplifies the 
quantum 
advantage to
\begin{equation}
    Q_A = 1+\frac{\xi}{N_S(1-\xi)}.
\end{equation}
In this form it is clear that in the $\xi/(1-\xi)\ll N_S$ setting, all the 
quantum advantage is destroyed. Conversely,
an unlimited quantum advantage can be observed in the $N_S\ll\xi$ regime.

\subsection{Receiver Operating Characteristic performance}

In this section, we analyze the quantum and classical noise radar in the 
asymmetric setting, i.e., when the prior probabilities are not equal, under a 
different perspective. We derive the ROC curves in the case when the idler and 
the signal are separately measured with heterodyne detection. 
In addition, we show the asymptotic performances of the optimal quantum 
strategies. 

\subsubsection{ROC curves with heterodyne detection}

\begin{figure}
    \centering
    \includegraphics[width=\linewidth]{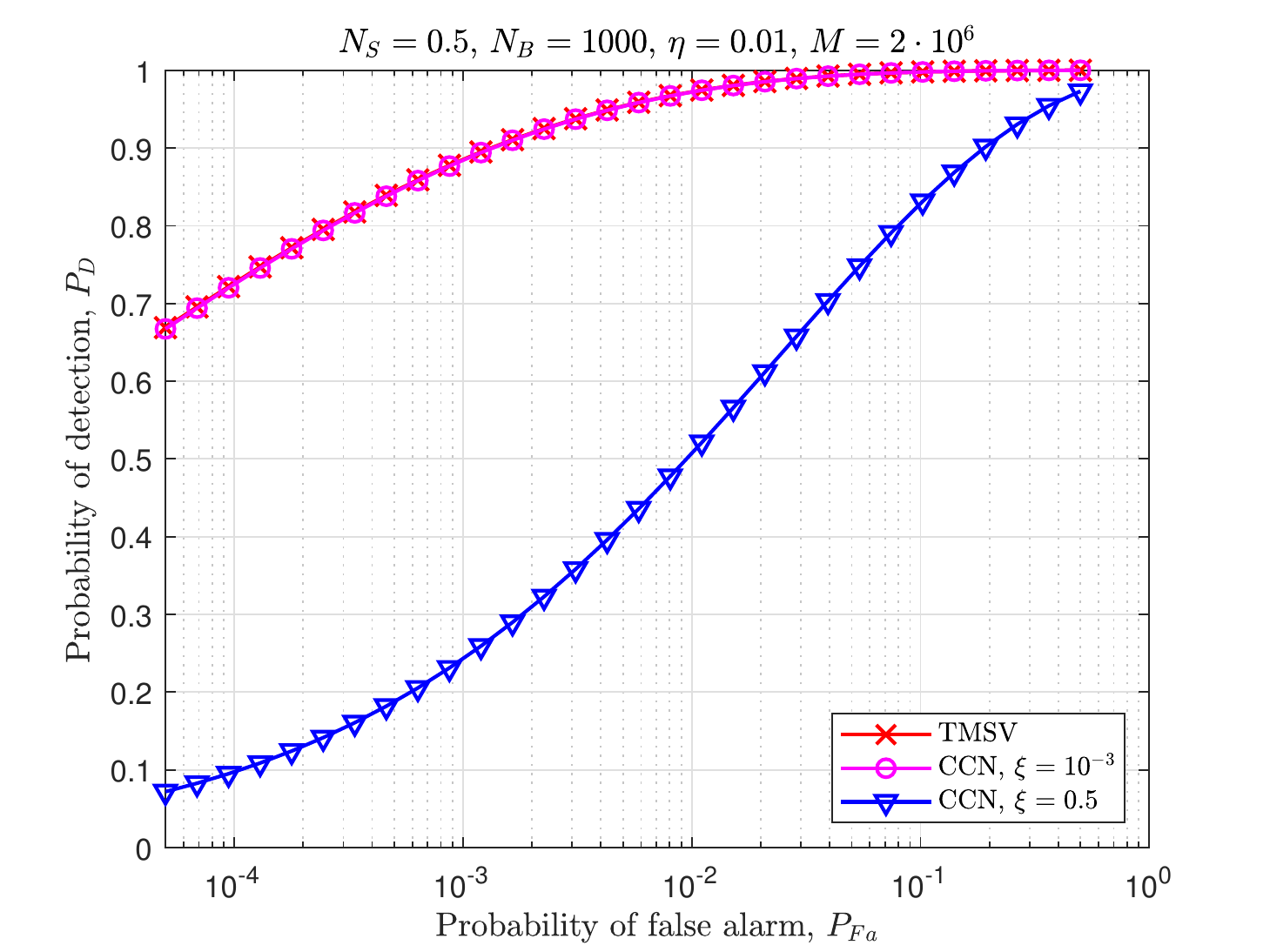}
    \caption{{\bf ROC curves with heterodyne detection for different radar 
    systems}. In the optimal classical setting, corresponding to 
    $\xi/(1-\xi)\ll N_S$, the classical noise radar has the same ROC curve as 
    the quantum noise radar. This shows how the entanglement initially present 
    in the system is not properly exploited when the received signal and the 
    idler are measured separately.}
    \label{fig:ROC}
\end{figure}

   We assume measurement is performed with heterodyne detection of the four 
   quadratures $\Vec{X}=\braket{\hat{X}}$. Based on the observed data set 
   $\mathbf{X}=\{\Vec{X}_1,\Vec{X}_2,\ldots,\Vec{X}_M\}$, an optimal threshold 
   detector   is computed by maximizing the log-likelihood ratio test around 
   $\eta\ll 1$. 
   This detector has, by Wilks's theorem, asymptotically Chi-squared 
   statistical distribution.     The ROC is computed in terms of the 
   probability of detection ($P_D=1-P_M$) 
    as 
    \begin{equation}
        P_D = Q_{\chi_1'^2(2M\kappa^2)}\left(Q_{\chi_1^2}^{-1}(P_{Fa})\right),
    \end{equation}
    where $Q$ is defined as the right-tail probability of the respective 
    distribution. The performance for this detector is presented with an 
    example in Fig.~\ref{fig:ROC}, for realistic parameters in the microwave 
    regime. We see that the quantum source is not any better than a classical 
    source, when $\xi/(1-\xi)\ll N_S$.
This is due to the amount of noise added with heterodyne detection. 

\subsubsection{Quantum Stein's lemma}

\begin{figure}
    \centering
    \includegraphics[width=\linewidth]{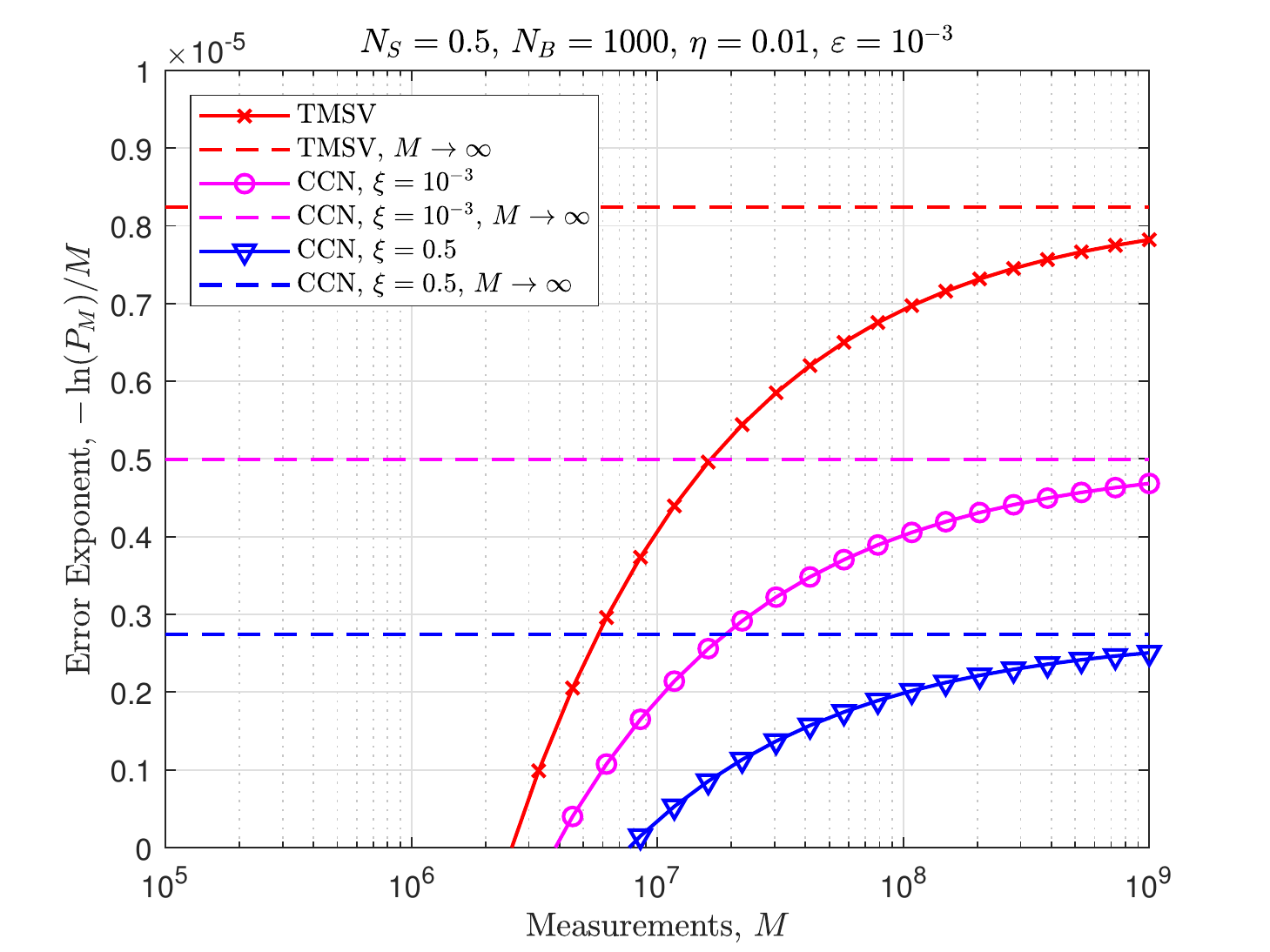}
    \caption{{\bf Asymptotic error probability exponent in the asymmetric 
    setting.} This is computed according to the quantum Stein's lemma (see 
    Eq.~\eqref{Steins}). Here, we consider an optimal joint measurements of the 
    idler and received signal for both the quantum and the classical radar 
    systems. The amount of measurements $M$ needed for the error probability 
    exponent to converge to the asymptotic values increases  logarithmically in 
    $\varepsilon^{-1}$, see Eq.~\eqref{Steins}. In the asymptotic regime, a 
    $\ln (1+1/N_S)$ advantage is achievable for $N_S\ll1$, see 
    Eqs.~\eqref{DQ}-\eqref{DC}. However, a substantial advantage is achievable 
    also for moderate values of $N_S$ (in the figure $N_S=0.5$).}
    \label{fig:errorexp}
\end{figure}

We provide the asymptotic performance  as given in 
Subsection~\ref{relative_entropy}. This analysis has been done for a radar 
system based on coherent state~\cite{Wilde2017}, but the classically-correlated 
noise case was missing. The quantum relative entropy rules the asymptotic 
behaviour of the error probability decay. These are given by
\begin{equation}\small
     D_\text{TMSV}\simeq\frac{\eta 
     N_S(N_S+1)}{N_S+N_B+1}\left[\ln\left(1+\tfrac{1}{N_B}\right)+\ln\left(1+\tfrac{1}{N_S}\right)\right],
      \label{DQ}
\end{equation}
\begin{equation}\small
   D_\text{CCN}\simeq\frac{\eta N_S^2}{N_S-\tfrac{\xi N_B}{1-\xi} 
   }\left[\ln\left(1{+}\tfrac{1}{N_B}\right){-}\ln\left(1{+}\tfrac{\xi}{N_S(1-\xi)}\right)\right],\label{DC}
\end{equation}
to  first order in $\eta$. One can easily see that the $\xi/(1-\xi)\ll N_S$ 
setting is optimal for the classical case. In particular, in the $N_B\gg1\gg 
N_S\gg\xi$ regime, we have that $D_\text{TMSV}/D_\text{CCN}\sim\ln(1+1/N_S)$. A 
substantial advantage can be also found for moderate values of $N_S$ (see 
Fig.~\ref{fig:errorexp}). The variance of the quantum relative entropy provides 
the convergence rate of the error probability exponent to its asymptotic value, 
for the type-I error probability constrained to be lower than $\varepsilon$. We 
do not provide the explicit formula here. However, in Fig.~\ref{fig:errorexp} 
we depict an example for the values of $\xi$ previously considered, and for 
$\varepsilon=10^{-3}$.  

\section{Conclusion and Outlook}\label{sec:conclude}

We have compared the performance of a quantum noise radar based on two-mode 
squeezed states and a class of noise radars based on thermal states. We have 
found that, given a constraint on the transmitting power, a quantum advantage 
in the ROC curve and their asymptotic performances is possible. 
If a low-powered signal of a quantum noise radar is amplified, then there is a 
classical noise radar which outperforms considerably the quantum radar. In 
addition, if enough noise is added at the idler level, such as when it is 
amplified or measured heterodyne, then all the quantum advantage is lost. 
However, a quantum advantage appears when the idler and signal are allowed to 
be measured jointly. Our results show that amplification is not a good strategy 
to overcome the technical difficulties that one must face in a practical 
quantum radar implementation. This suggests that quantum radars are more 
difficult to achieve than what recent experiments were 
claiming~\cite{barzanjeh2019experimental, chang2019quantum}. Interfacing a 
large bandwidth entangled microwave signal with the environment, and developing 
low-noise power detectors is crucial for performing a QI experiment with a 
quantum advantage.

\section*{Acknowledgment}

The authors would like to thank Per Delsing and Philip Krantz for very useful 
discussions, and Jeffrey H. Shapiro for reviewing the manuscript.  The authors 
acknowledge the Knut and Alice Wallenberg foundation for funding. RDC 
acknowledges support from the
Marie Sk{\l}odowska Curie fellowship number 891517 (MSC-IF
Green-MIQUEC).

\bibliography{bibi}

\end{document}